\documentclass[aps,prl,twocolumn,showpacs,preprintnumbers,superscriptaddress,floatfix,amsmath,amssymb,amsfonts,10pt,bibliography]{revtex4-1}
\usepackage[utf8]{inputenc}
\usepackage{amsmath,amsfonts,amssymb}
\usepackage{graphicx}
\usepackage{color}
\usepackage[sf]{subfigure}
\usepackage{fouriernc}
\usepackage{xcolor}
\usepackage{bm}
\usepackage{dcolumn}

\usepackage[colorlinks=true,
linkcolor=blue,
urlcolor=blue,
citecolor=blue,
unicode,
pdfencoding=auto]{hyperref}
\usepackage{etoolbox}

\begin{document}

\title{Energetics of the PH-Pfaffian state and the 5/2-fractional quantum Hall effect}
\author{Edward H. Rezayi}
\affiliation{Department of Physics, California State University Los Angeles, Los Angeles, California 90032, USA}
\author{Kiryl Pakrouski}
\affiliation{Department of Physics, Princeton University, Princeton, New Jersey, USA}
\author{F. D. M. Haldane}
\affiliation{Department of Physics, Princeton University, Princeton, New Jersey, USA}
\begin{abstract}
{We present a method for the  exact construction of the fully particle-hole symmetric PH-Pfaffian 
ground state and its charged
excitations on a sphere.  We adopt the Moore-Read state, but with a
non-holomorphic pairing component as in previous studies, and project it to the lowest Landau level. 
We study the energetics as well as other properties of these states and find that in a pure system 
interacting with the Coulomb forces the PH-Pfaffian
cannot compete with either the Moore-Read state or its particle-hole conjugate, 
the anti-Pfaffian state, as an explanation for the 5/2-effect. 
}

\end{abstract}

\maketitle

One of the most intriguing topological\cite{Wen1990,*Wen1992} quantum phases of matter was 
discovered\cite{Willett1987} in the 
fractional quantum Hall effect (FQHE)\cite{RevModPhys-Nayak-Simon-2008,RevModPhys-Hansson-Simon-2017} at 5/2 filling of the lowest two Landau levels (LLs). 
A large
number of studies of the 5/2-state point to either the Moore-Read\cite{MR1991nonabelions} Pfaffian (MR-Pf)
state, or it's particle-hole (PH) conjugate, the anti-Pfaffian (aPf) state\cite{Levin-Apf,Nayak-Apf} 
to explain this phenomenon.
Another related state that has recently attracted considerable attention is the 
PH-Pfaffian\cite{SonPRX.5.031027} (PH-Pf). It is so named because, unlike the MR or the aPf, this state is symmetric under PH conjugation.  All three are expected to be Hall superconductors\cite{Read-Green-2000}, but with different pairing symmetries. 
There is however
scant support in numerical studies of the 5/2-state for the PH-Pfaffian. Instead, there is considerable 
evidence in favor of the  MR and aPf. Some examples in different geometries are given in these papers:  
\cite{Morf1998-spin,
Rezayi-Haldane2000,feiguin-DMRG-2008,Peterson-Jolicoeur-2008,XinWan-etal-disk-2008,
Sheng-Haldane-2011}.

Earlier studies, for the most part, preserved the P-H symmetry of the Hamiltonian 
and were unable to discriminate between the 
latter two 
ground states.  In the presence of inter-Landau-level transitions or mixing (a  ubiquitous feature of experiments), PH symmetry is 
broken  and the aPf gains the advantage\cite{Simon-Rezayi-2013,*Rezayi-simon-2011,Zaletel-etal-2015,Rezayi-2017}.  However, the energy splitting per particle is small
and 
omitting some pseudopotential components of the 3-body 
mixing\cite{Sodemann-2013} corrections stabilizes\footnote{If the first 5 as opposed to 6 or 
more 3-body pseudopotentials are included\cite{Pakrouski-etal-2020}}
the MR-Pf\cite{Pakrouski-etal-2016,Peterson-Nayak-2013,Jain-Toke-Wojs-2010}.
The quasiparticle excitations of all three states possess Majorana zero modes and are expected to obey
non-Abelian statistics\cite{MR1991nonabelions,Bonderson-etal-2011}, which is a necessary ingredient for quantum information processing. They are also fully spin-polarized, in agreement with both experiment\cite{Gamez2012-spin,Eisenstein2017-spin} and numerical 
calculations\cite{Morf1998-spin,feiguin2009-spin} of the 5/2-effect. 

Recent measurements\cite{Banerjee-etal-2018} of quantized thermal Hall conductance $\kappa_{xy}$, however, found  a value that is only consistent with
the PH-Pf state.  There are several interesting scenarios for explaining this  observation. 
Disorder, which is present in experiment, has been put forward as the decisive factor
in stabilizing the PH-Pfaffian\cite{Feldman2016}. 
Another  
possibility is the formation of Pf and aPf 
domains in the presence of disorder\cite{Wang-etal-2018,Mross-etal-2018,Lian-Wang-2018}, 
which under suitable conditions
could result in the measured quantized thermal Hall 
conductance.  Whether this mechanism can account for the experimental observation is 
unclear\cite{Simon-etal-pf-apf-2020,zhu2020top}.

Another possibility is that the aPf ground state, under certain 
conditions could produce the measured $\kappa_{xy}$\cite{Simon-Rosenow_2020,Mulligan_2020}. 
However, a more recent experiment also supports a PH-PF ground state\cite{dutta2021}.
In any event, these developments call for a thorough examination of the PH-Pfaffian state.

In this paper we formulate an exact
procedure for calculating the ground state and charged quasiparticle excited states of the PH-Pfaffian.
We then obtain 
results for up to fourteen and twelve electron systems for the ground and charged excited states 
respectively.  These sizes are comparable to previous exact diagonalization studies of the 5/2-effect. 
We use the spherical geometry since the angular momentum ``technology'' 
simplifies the construction.  In what follows, all energies are
given in units of $e^2/4\pi\epsilon\ell_B$. Distances (wavevectors) are given in units of the 
magnetic length $\ell_B$ (inverse magnetic length) and densities in inverse $2\pi\ell_B^2$ units.
\begin{figure}[t]
\centering
{\includegraphics[height=2.8in,width=3.6in]{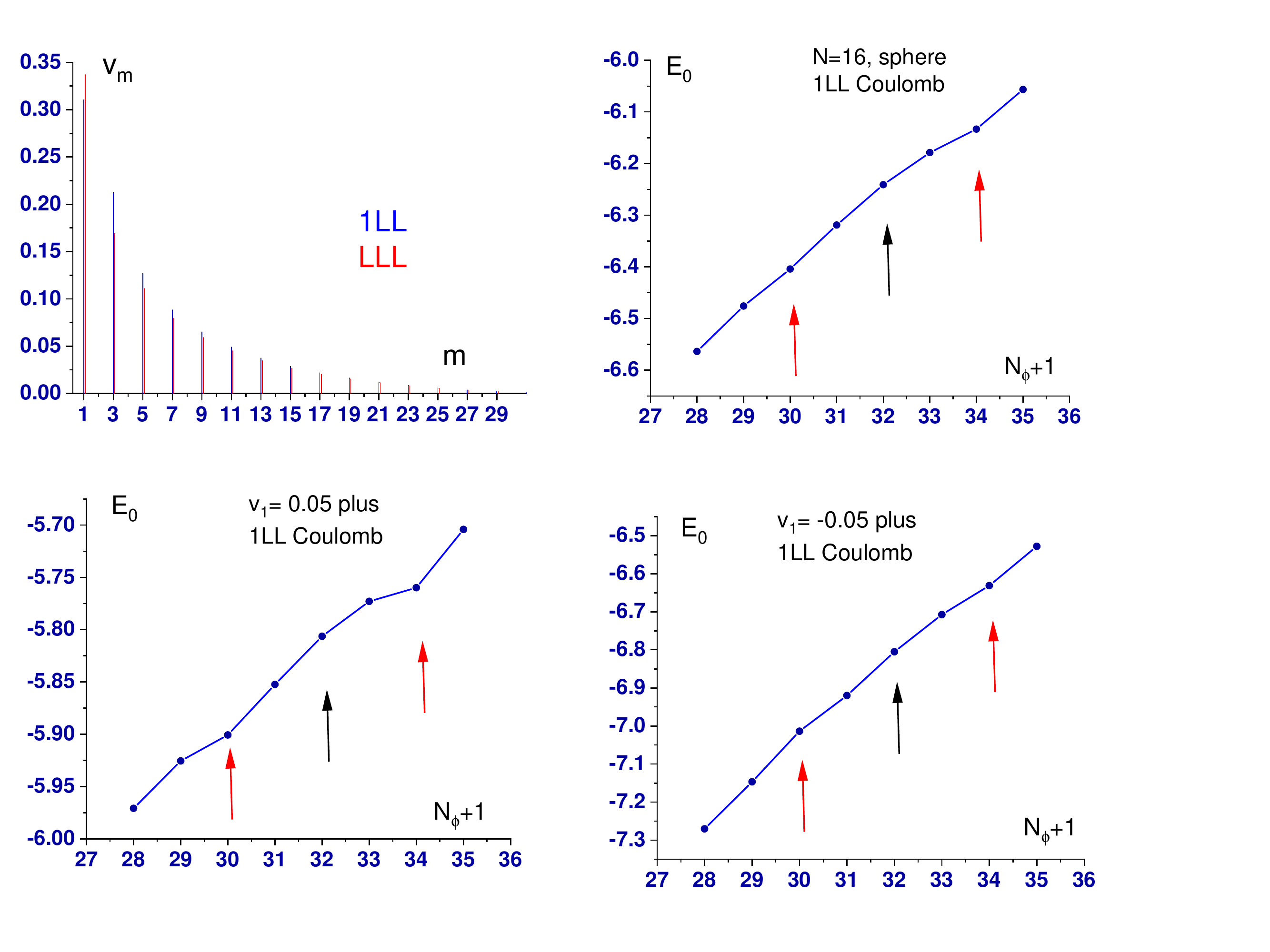}}
\caption{\label{fig:SLLgaps}%
Top left: in first row: comparison of pseudopotentials $v_m$, for spin-polarized electrons  of the 
lowest and first excited  Landau levels. Top right: ground state energies for the Coulomb potential in
1LL  as the flux $N_\phi$ is varied 
for 16 electrons. Bottom: $v_1$ of 1LL Coulomb is varied by $\pm 0.05$.}
\end{figure}
As a reminder, Fig. \ref{fig:SLLgaps}  shows  the gapped phases of FQHE near the 
half-filled first excited Landau level (1LL)  as the first Haldane pseudopotential $v_1$ is varied. 
Only the first 3 
odd pseudopotentials seem substantially different from their lowest Landau level (LLL) values.  
The arrows show 
the positions of the MR, PH-Pfaffian, and the anti-Pfaffian from left to right respectively.  
The only visibly gapped states 
appear to be MR-Pf and aPf, which are related by particle-hole conjugation that maps the 
electrons to holes and vice versa.  The MR-Pf satisfies the relation $N_\phi=2N_e-3$, which 
has a shift of 3.  The shifts of the PH-Pf and aPf are 1 and -1 respectively.  Different
shifts generally signify a different topological phase of matter.

{\it Computation of Wavefunction}- While there may well be other forms for 
the PH-Pf wavefunction we will use the one from previous studies\cite{Feldman2016,Mishmash-etal-2018,Balram-etal-2018,Mross-2020}:

\begin{equation}
\lvert\Psi_{\text{PH-Pf}}(\{{\bm r_i}\})\rangle=
\text{Pf}_{i,j}\left\{ \frac{1}{\bar{u}_i\bar{v}_j-\bar{u}_j\bar{v}_i}\right\}
\lvert\Psi_{1/2}\rangle,
\end{equation}
where $u$ and $v$ are spinor coordinates\cite{Haldane-1983} and the holomorphic part $|\Psi_{1/2}\rangle$ is the $\nu=1/2$ bosonic Laughlin state:
\begin{equation}
\lvert\Psi_{1/2}(\{ u_i,v_i\})\rangle=\prod_{i>j}(u_iv_j-u_jv_i)^2.
\end{equation}
Projection of the wavefunction to the LLL turns $\bar{u}$ and $\bar{v}$ into operators (usually derivatives\cite{Jain-Kamilla-1997}). 
The key idea in our approach is to project one pair at a time. The projection operators are only in 
the Pfaffian, which is the pairing part of the wavefunction.
We thus start with projecting a single pair. Multiplying both numerator and denominator by the 
factor $u_iv_j-u_jv_i$, we have: 
\begin{equation}
\frac{1}{\bar{u}_i\bar{v}_j-\bar{u}_j\bar{v}_i}=\frac{u_iv_j-u_jv_i}{|u_iv_j-u_jv_i|^2}.
\end{equation}
This is a rotationally invariant holomorphic pair (scalar) operator with a $1/r^2$ potential, 
where $r$ is the chord
distance between particles $i$ and $j$ on a unit sphere. This potential is to be projected into the LLL.
The numerator is holomorphic and turns a 2-boson state into a state of  two fermions, which in total 
adds a flux quantum $N_\phi^{F}=N_\phi^{B}+1$, without altering $J$ and $M$:
$\lvert J,M,N^B_{\phi} \rangle$ transforms to  $\lvert J, M,N^F_{\phi} \rangle$. $J$ and $M$ 
are the total and azimuthal
angular momenta of the pair. Using the Wigner-Eckart theorem we obtain  the reduced matrix 
elements below. These are, in fact, the Haldane pseudopotentials for a $1/r^2$ ``Hamiltonian'' that changes a pair of bosons into a pair of fermions. Therefore, we set  $M=J$, simplifying the 2-particle 
wavefunctions\cite{Haldane-1983} to: 

\begin{equation}
\lvert J,J;N^{B(F)}_\phi\rangle=(u_iv_j-u_jv_i)^{N^{B(F)}_{\phi}-J}u_i^Ju_j^J, 
\end{equation}

where $B(F)$ refers to bosons(fermions). The number of bosons and
fermions are equal and is denoted by $N_e$.
The matrix element of the pair-operator between the 2-particle states is reduced 
to the expectation value of $1/r^2$ (apart from normalization factors) for a two-fermion state. 
The pseudopotentials are:

\begin{equation}
V_J=\frac{N^F_{\phi}+1}{\sqrt{(N^F_{\phi}-J)(N^F_{\phi}+1+J)}}
\end{equation}

To get the matrix elements in a more convenient form (Eq.~(\ref{Eq-ME})), we expand the pair creation 
annihilation operators in terms of a pair of single particle boson annihilation and a pair of fermion creation
operators. Again, because of the additional flux quantum for fermions relative to bosons,
the needed Clebcsh-Gordan (CG) coefficients for the same 
$J$ and $M$ have the correct parity under particle exchange for both bosons and fermions. Combining the CG 
coefficients with $V_J$ and summing over $J$ and $M$ yields the desired matrix elements,
which can be separately calculated and stored:

\begin{align}
V(m_i^f,m_j^f;m_i^b,m_j^b)&=\langle m_i^f,m_j^f\lvert\frac{u_iv_j-u_jv_i}{|u_iv_j-u_jv_i|^2}\rvert m_i^b,m_j^b\rangle, \nonumber  \\
m_i^f+m_j^f&=m_i^b+m_j^b \label{Eq-ME}.
\end{align}
The matrix elements can easily be antisymmetrized in the two fermion and symmetrized in 
the two boson orbitals.

The coordinates in the Pfaffian can now be integrated out.   The antisymmetrization required 
in the Pfaffian can, by a change of integration variables, be compensated by the exchange of 
fermion orbitals.  
The inter-pair anti-symmetrization of the fermion orbitals only requires 
$N_{fact}=N_e!/(2^{N_e/2}(N_e/2)!)=(N_e-1)!!$ independent terms, which is much smaller than N!. 
However, this operation has to be done for all occupied single particle states with total 
zero azimuthal 
angular momentum. The total number of configurations for fermions is $Nc = N_{fact} N_H$,
where $N_H$ is the dimension of the appropriate many-body fermion Hilbert space\footnote{The relevant 
information on fermions can be calculated separately and stored (in a peicemeal manner if necessary).  Several
of the fermion occupations, permutation signs and hashtag addresses can be stored within integers}.

\begin{table}
\caption{\label{PH-Pf-MR-Pf} Some attributes, indicated by the column headings, of PH-Pf and MR states for different sizes $N_e$.}
\begin{tabular}{r|c|c|c|c}
$N_e $&$ \vert \langle \Psi\vert \Psi_{\text{Sym}}\rangle\vert$&Variational E$_0$&E$_0/N_e$ & E$_0$(Pf)$/N_e$ \\ \hline\hline
6$\ $& 0.9999996 & -2.583729 &$\ $ -0.4306215 & -0.4868794 \\
8$\ $& 0.9999633 & -3.291081 &$\ $ -0.4113851 & -0.4458210 \\
10$\ $& 0.9999807 & -3.993417 &$\ $ -0.3993417 & -0.4248679 \\
12$\ $& 0.9999463 & -4.694213 &$\ $ -0.3911844 & -0.4122298 \\
14$\ $& 0.9998940 & -5.404673 &$\ $ -0.3860481 & -0.4040570 \\
\end{tabular}
\end{table}

The main calculation is organized in a {\em single} loop of size $Nc$ for fermions.
Because of the conseveration law for each pair of bosons in Eq.~(\ref{Eq-ME}), there are an additional 
$N_e/2$ inner loops for boson orbitals.
In the inner core of these $N_e/2+1$ loops the PH-PF wavefunction is obtained from the product of
the matrix elements, other information on fermion basis, and the Laughlin wavefunction.
be separately zero. 

While the code is very short and relatively simple, it still is an $N_e$-body 
operator with a much higher degree of complexity than diagonalizing a many-body Hamiltonian. On the
other hand, 
the computations 
for different sets of fermion orbitals $\{m_i\}$ are independent and the outer loop can be massively 
parallelized.  We have also taken advantage of reflection symmetry to divide the 
basis (by its parity), and hence the outer fermion loop,
into two independent, but nearly equal parts, providing further parallelization.

\begin{figure}[t]
\centering
{\includegraphics[height=1.5in,width=2.0in]{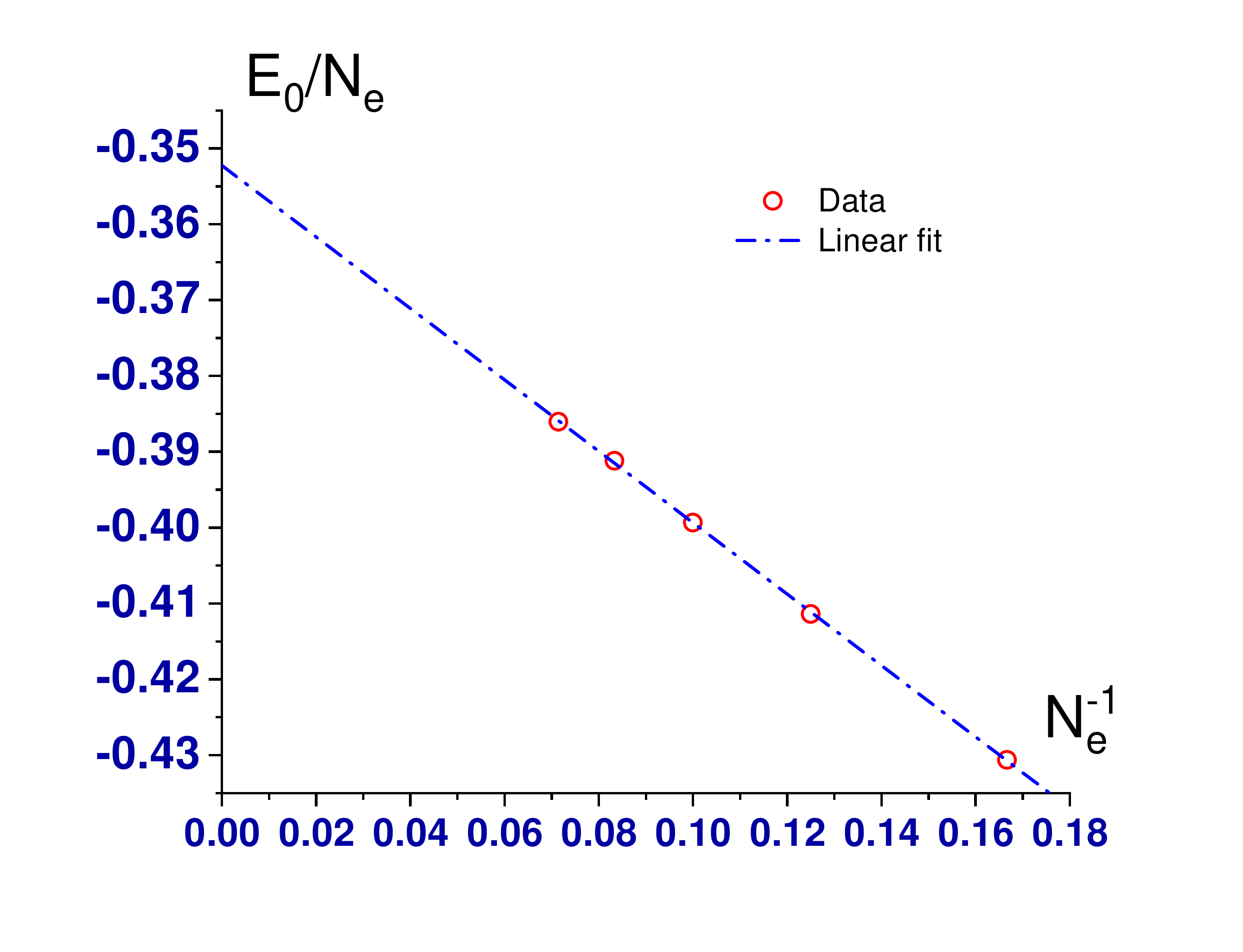}}
\caption{\label{fig:VarE-PHPF}%
Variational energies of the PH-Pfaffian state for 6-14 electrons. The straight line is a least 
squares fit of the data, yielding a an infinite-size value of $-0.3523\pm0.0004$ per electron.
}
\end{figure}

\begin{figure}[t]
\centering
{\includegraphics[height=1.5in,width=2.0in]{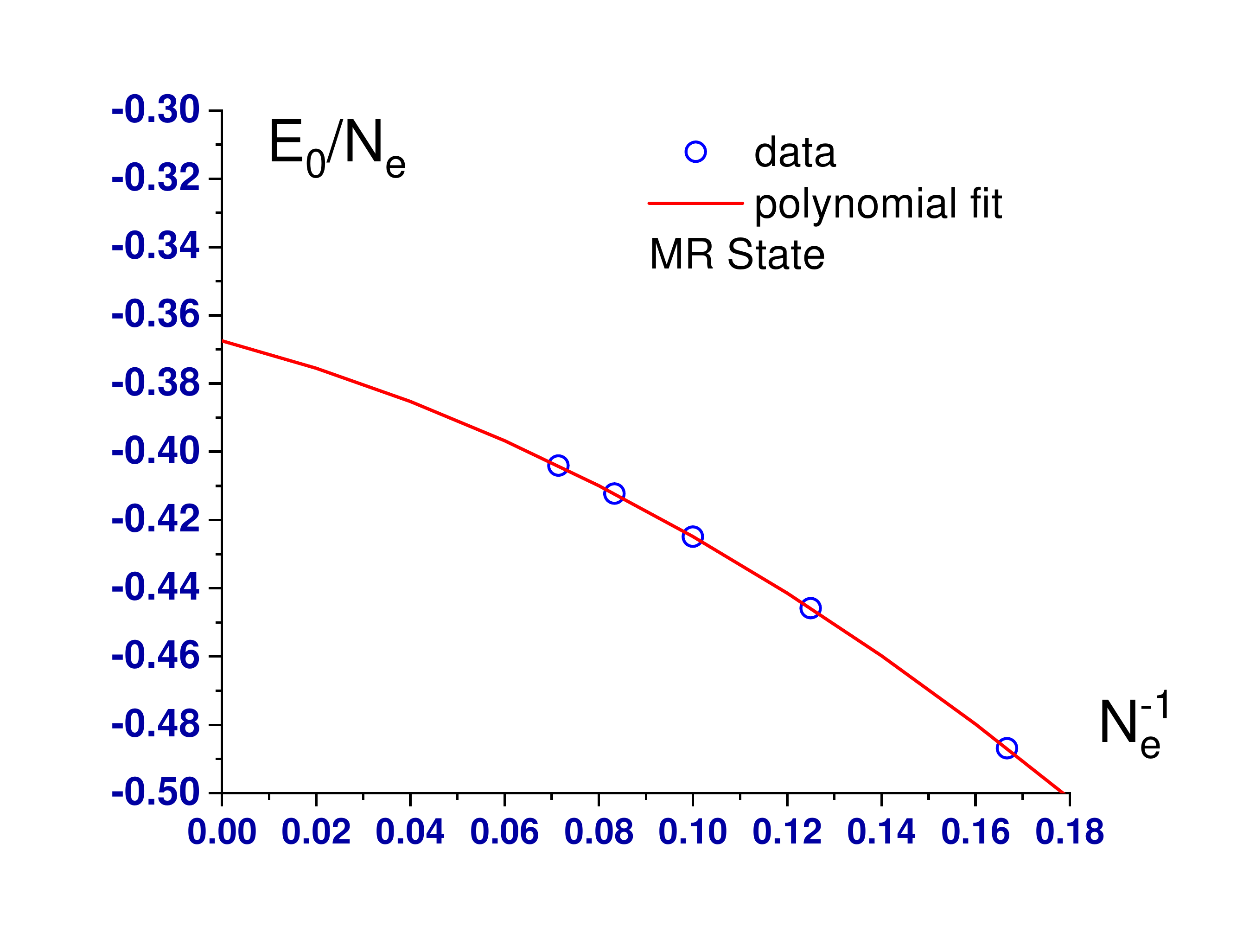}}
\caption{\label{fig:VarE-MR}%
Variational energies of the MR-Pfaffian state for 6-14 electrons. The curve is a 
fit of the data to polynomial of degree 2. The intercept in the  infinite-size limit is $-0.3675\pm0.0004$ per electron.
}
\end{figure}

{\it The Ground State (GS)}- The PH-Pfaffian wavefunctions are very nearly particle-hole symmetric. 
However, they cannot be fully symmetrized or anti-symmetrized by the usual means (making a linear combination of the two states) because of the anti-unitarity of the
PH-transformation. The problem is overcome if the eigenvectors of the $2\times 2$  overlap matrix of the two 
states related by PH-conjugation are obtained. The parity of the state is immaterial. One eigenvector would have an overlap of near 
unity with the calculated wavefunction (see Table \ref{PH-Pf-MR-Pf})  and the other a very small overlap.
The table  also shows their variational energies
for the 1LL Coulomb potential.  These are plotted in Fig. \ref{fig:VarE-PHPF} and give an extrapolation to infinite size of
$-0.3523$.  That is larger than $-0.3675$ for the Pf (or equivalently for aPf) energies extrapolated in Fig. \ref{fig:VarE-MR}.  
We note that the PH-Pf on the sphere is aliased (same $N_e$ and $N_\phi$) with
the particle-hole symmetric version of Jain's\cite{Jain-CF-1989} composite fermion (CF) with an
effective magnetic flux quantum of one: $N_\phi^*=N_\phi-2(N-1)=1$ as opposed to 
zero \cite{halperin_theory_1993,RR-CF-1994}.
This  has been called the Dirac CF (DCF) \cite{SonPRX.5.031027,Geraedts-Zaletel-Mong-2016} since its Berry phase, when taken around the Fermi 
surface, is $\pi$\cite{SonPRX.5.031027,Geraedts-Wang-2018,Wang-Geraedts-2019}.
Both composite Fermi liquids of CF and DCF are appropriate ground states 
in the LLL at $\nu=1/2$ but not at 5/2 filling. In the PH-symmetric 
case  the electrons form 
closed shells with total angular momentum $L=0$ for sizes given by $N_e=(n+1)(n+2)$, with  $n$ a non-negative integer. For partially filled shells, the inter-DCF distances can be maximized for non-zero
values of angular momentum, which vary systematically with size\cite{RR-CF-1994}.

\begin{figure}[b]
\centering
{\includegraphics[height=1.5in,width=2.0in]{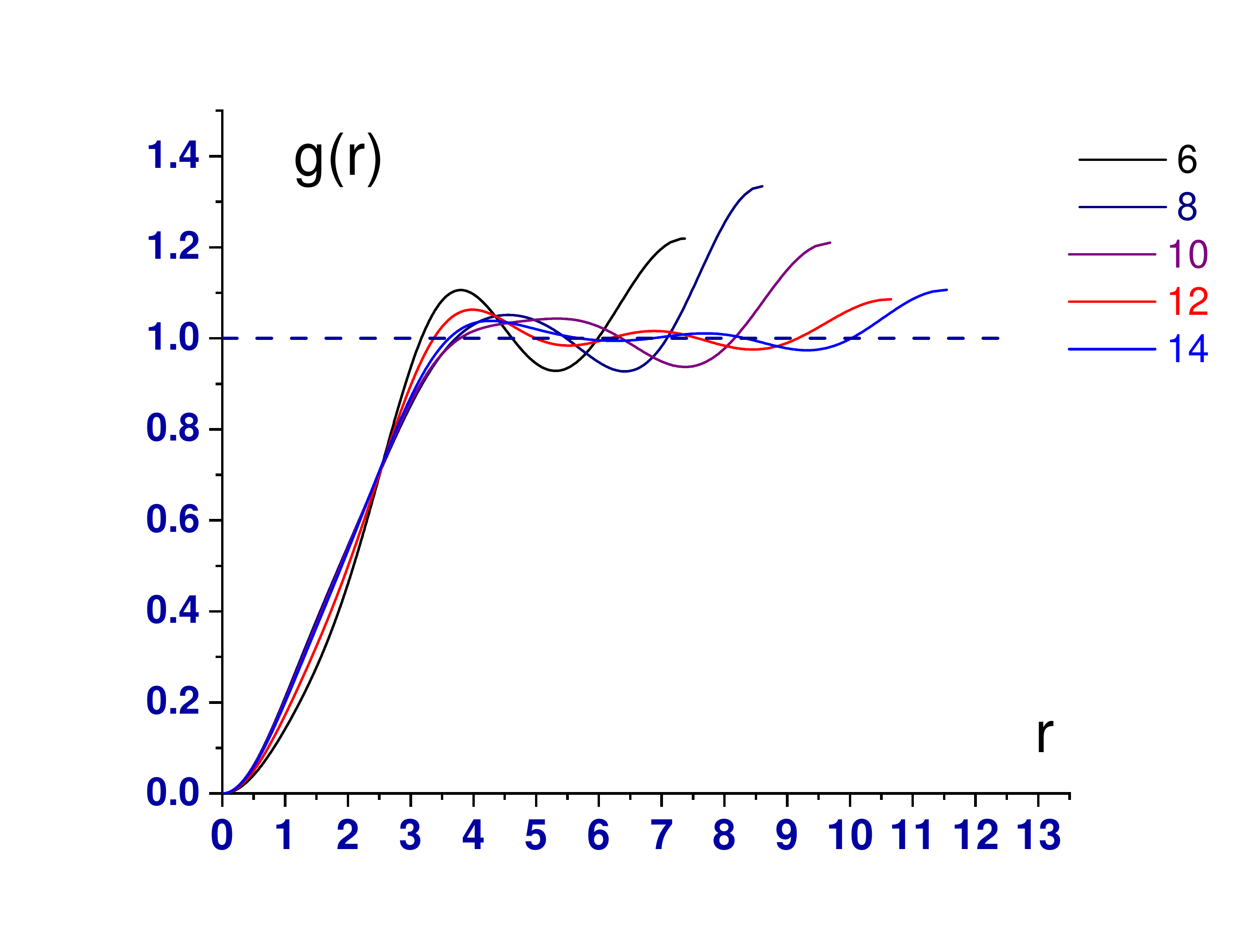}}
\caption{\label{fig:Pair-CorrF}%
Pair correlation function for 6-14 electrons as a function of the large circle distance.
}
\end{figure}

\begin{figure}[t]
\centering
\subfigure[\label{fig:S0Q_PH-Pf_fS}]{\includegraphics[height=1.25in]{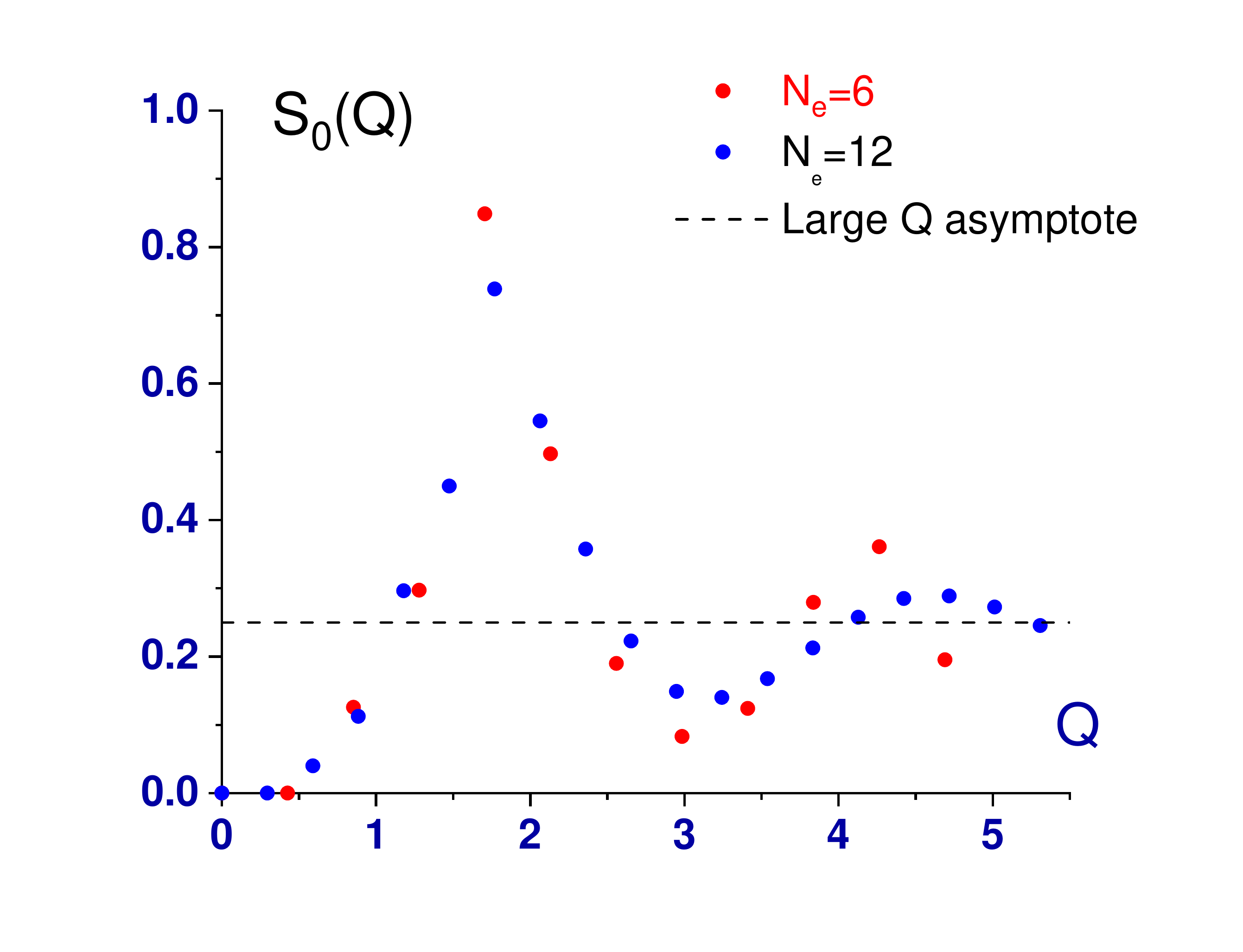}}
\subfigure[\label{fig:S0Q_PH_PF_NfS}]{\includegraphics[height=1.25in]{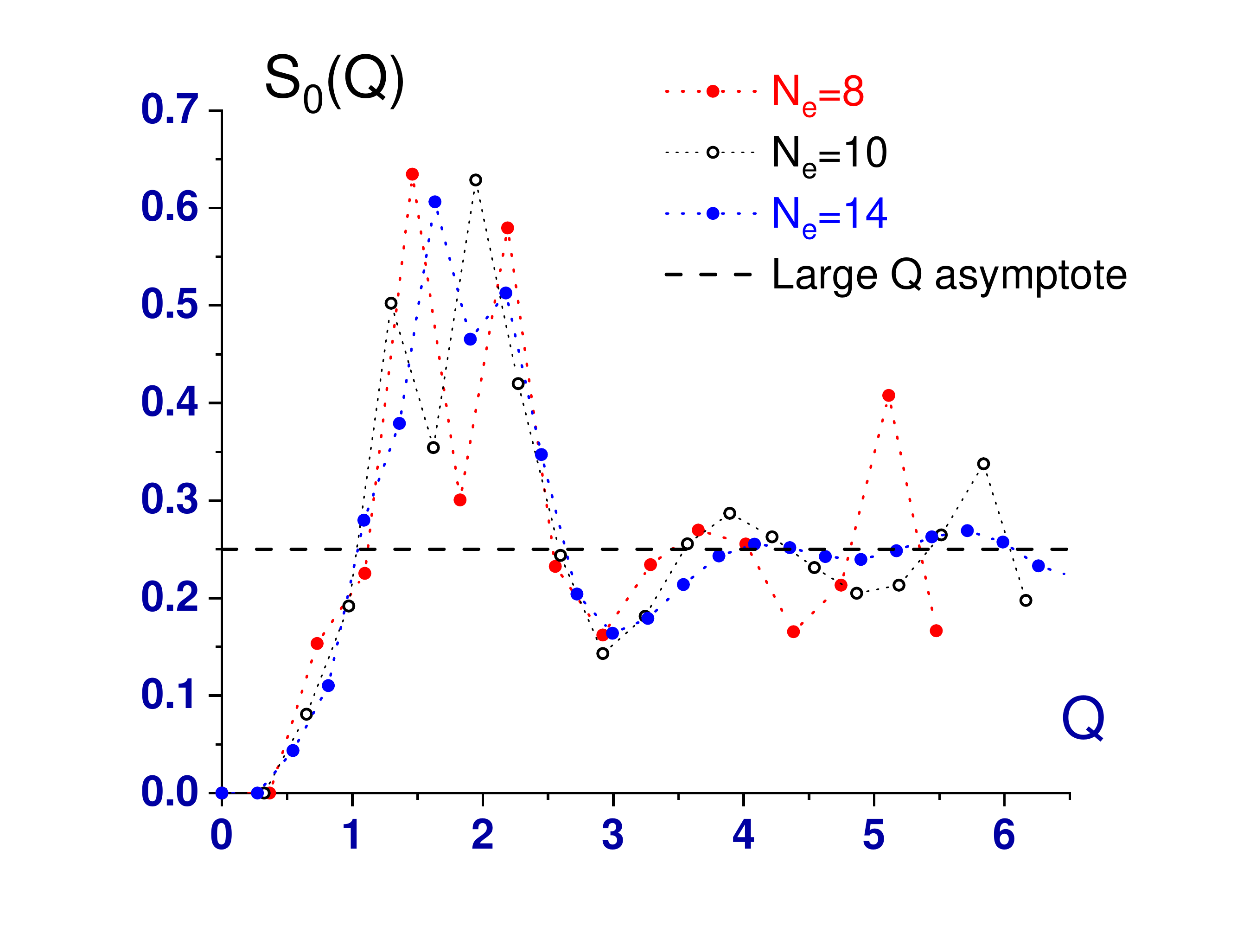}}
\caption{\label{fig:S0Q}%
(a) Guiding center structure factor with a single main peak for $N_e=6$ and 12.
(b) Same as in (a) except with two main peaks for $N_e=8$, 10, and 14. Filled (open) symbols are for cases
where the larger peak is to the left (right) of the other main peak.  The dotted line is the known 
asymptotic value\cite{*[{In this version the normalization of $S_0$ has been changed from $N_e$ to 
$N_\phi.\ \ $}] [{  }]  haldane_quantum_1990} of $S_0(Q)$ for large wavevector $Q$. 
}
\end{figure}

Fig. \ref{fig:Pair-CorrF} shows the pair correlation function for even sizes of 6-14 electrons.  Oddly,
there is no indication of convergence, in sharp contrast to the case of the MR-Pf state, where a clear 
picture emerges with 12 electrons\cite{RR-Pf-quasiholes}. In addition long-range tail oscillations, which are typical of composite Fermi liquids\cite{RR-CF-1994} persist to large sizes\cite{Mross-2020}.
Also, there appears 
to be two classes of states determined by whether the DCF's form a closed shell
or not.  Fig. \ref{fig:S0Q}  shows the (LL-independent) guiding center structure factor $S_0(Q)$.  We separate the
filled shell configurations $N=6$ and 12 ($n=1$, and 2 respectively) from the rest.  Only the first group exhibits a single sharp peak at a wavevector that approaches $2k_f$ for large sizes. 
This separation agrees with the high overlap of DCF with the PH-Pf for $N=12$ obtained by the 
Monte-Carlo method\cite{Mishmash-etal-2018}.
Since the PH-Pf is in fact a paired state of DCF's, this trend is not entirely surprising.  
For unfilled shells
the angular momentum of DCF is non-zero and, thus,  will have no overlap with the PH-Pf GS.   
However, these  trends may not bode well for a gapped topological phase. Moreover, 
the n=1 LL Coulomb potential 
is insufficient for the pairing of DCF's into a Hall superconductor and it is left as a compressible state. It seems unlikely that disorder can overcome these shortcomings.

{\it Charge Excitations}-To complete the picture of the PH-Pf, we turn to the quasielectron 
and quasihole excitations.  These, given below, are the most natural extension of the 
ground state wavefunction:

\begin{equation}
|\Psi_{\text{QE}}({\bm r_i})\rangle=
\text{Pf}_{i,j}\left\{ \frac{\bar{u}_i\bar{v}_j+\bar{u}_j\bar{v}_i}{\bar{u}_i\bar{v}_j-\bar{u}_j\bar{v}_i}\right\} \label{Eq-QE}
|\Psi_{1/2}\rangle,
\end{equation}
and
\begin{equation}
|\Psi_{\text{QH}}({\bm r_i})\rangle=
\text{Pf}_{i,j}\left\{ \frac{{u}_i{v}_j+{u}_j{v}_i}{\bar{u}_i\bar{v}_j-\bar{u}_j\bar{v}_i}\right\}
\label{Eq-QH}
|\Psi_{1/2}\rangle.
\end{equation}

\begin{figure}[t]
\centering
{\includegraphics[height=2.5in,width=3.0in]{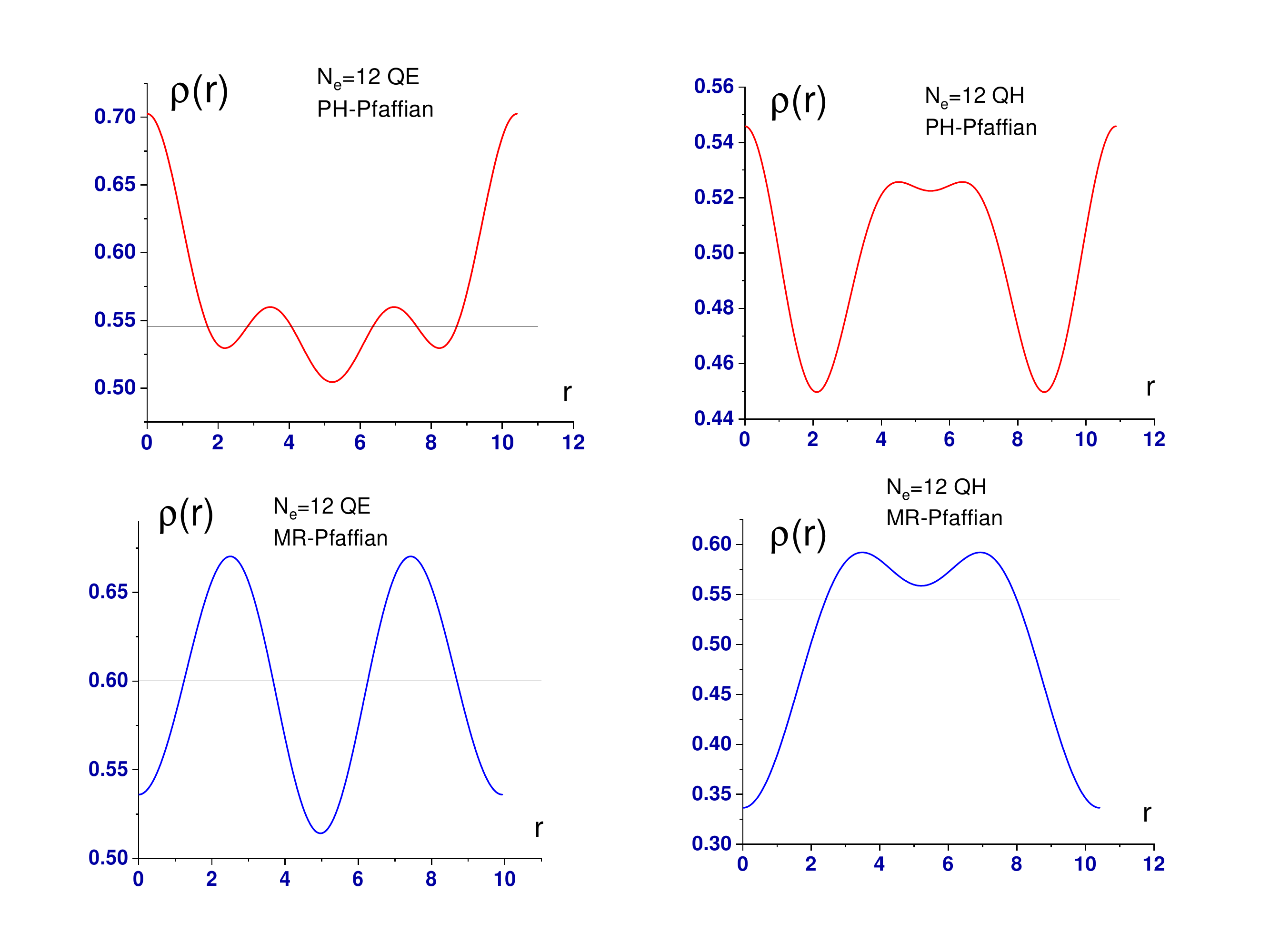}}
\caption{\label{fig:QP-QH-densities}%
The density of the PH-Pf (upper figures) two quasielectrons  (QEs) and two quasiholes (QHs) 
for 12 electrons  as a function of 
the large circle distances. The horizontal lines mark the density of the fluid  if the charge was  
distributed uniformly. The lower two figures are the MR QEs and QHs densities.
}
\end{figure}

\begin{table}
\caption{\label{VarE-Ph-Pf-MR}The variational energies of GS, charge excitations, and the gaps of 
	PH-Pf and
		MR state for 12 electrons.}
		\begin{tabular}{r|c|c|c|c}
		State & QE Energy & QH Energy & GS Energy& $\Delta_c$ \\ \hline\hline
		PH-Pf&-4.774736&-4.615954 &-4.694213  & -0.001132 \\
			MR-Pf&-5.028995 & -4.833692 & -4.946758 & 0.015415 \\
			\end{tabular}
			\end{table}

\begin{figure}[b]
\centering
{\includegraphics[height=1.5in,width=2.0in]{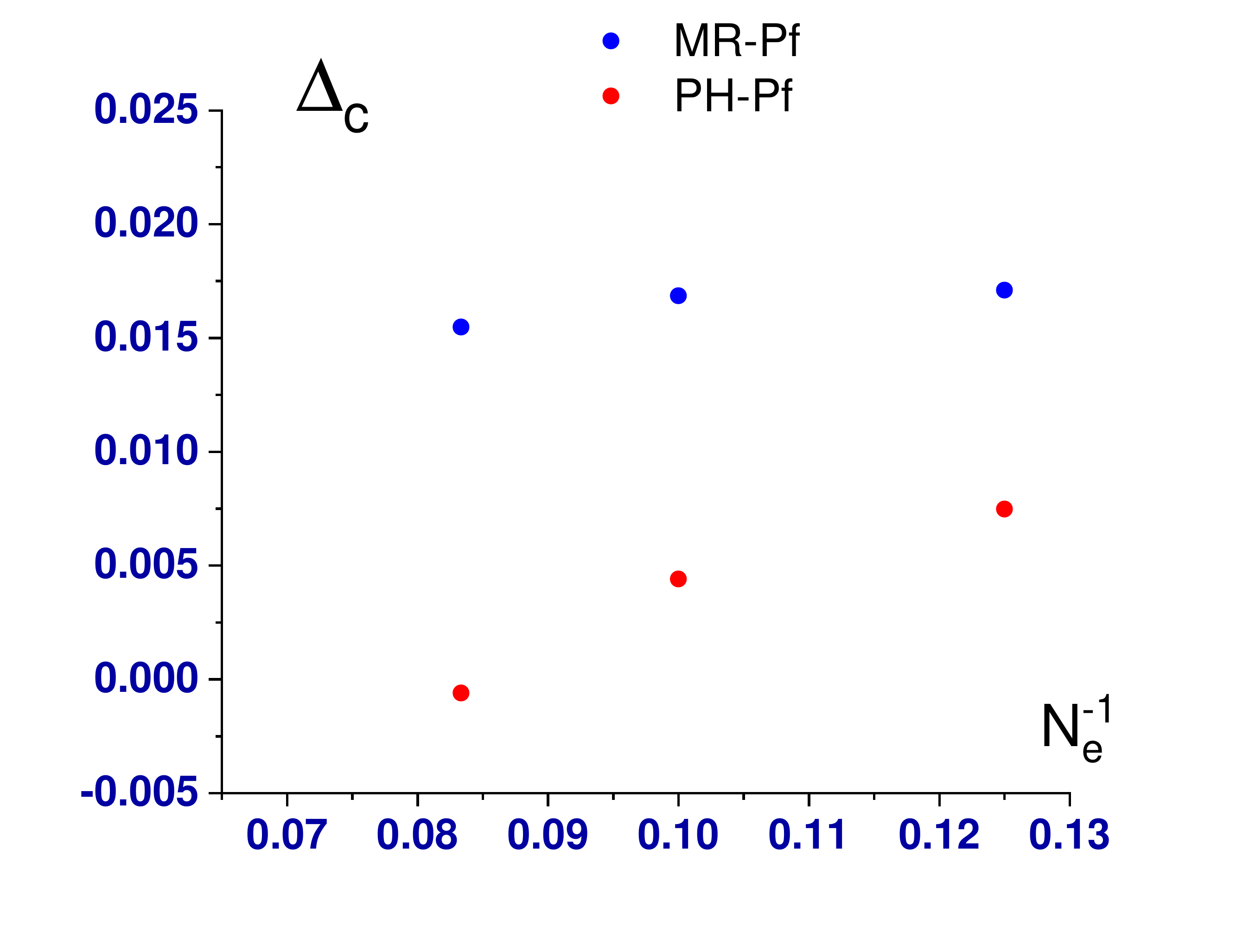}}
\caption{\label{fig:Gaps-MR-PHPf}%
Charge gaps in  8-12 electron systems for both MR and PH-Pf states.
}
\end{figure}

The two quasiparticles are at the poles of the sphere. As a result, the full rotational symmetry is 
downgraded to azmuthal symmetry.  We have included the corresponding MR quasiparticle 
states for comparison.  The wavefunction of a pair of quasielectrons and quasiholes\cite{RR-Pf-quasiholes,XinWan-etal-disk-2008} are the same as in Eqs.~(\ref{Eq-QE}) and (\ref{Eq-QH}), but 
with a holomorphic denominators\footnote{The spherical version of the wavefunction on 
the disk can be obtained by stereographic projection. The two poles are mapped to 
infinity (north pole) and to the origin. In the notation of Ref. \cite{RR-Pf-quasiholes} 
the (complex) positions of the quasiholes $ w_1\rightarrow 0$ and $ w_2$ becomes an overall factor,
which can be dropped. This yields the form\cite{XinWan-etal-disk-2008} used in this paper.}.
The calculation becomes a little more complicated, due to the loss of full rotational
symmetry, but the
matrix elements of the 2-body interactions can still be computed by transformation of coordinates 
\cite{Fano-etal-1986}. Fig. \ref{fig:QP-QH-densities} shows the densities of these states for $N=12$.  We also compare the variational 
energies in Table \ref{VarE-Ph-Pf-MR}. 
Again, the MR state and the corresponding quasiparticles have lower energies. Whether they remain so in the thermodynamic limit is unclear. 
A more meaningful comparison would be to calculate the gaps for  creating a neutral pair of 
quasiparticles. Since a pair of quasiparticles is created for each quantum of flux 
above or  below the GS, we divide the  energies by two. The gap for creating the neutral pair is
$\Delta_c$ defined by: 
\begin{equation}
\Delta_c=\frac{E(N_{\phi}+1,N_e)+E(N_{\phi}-1,N_e)-2E(N_{\phi},N_e)}{2},
\end{equation}
where $N_\phi$ is the number of flux quanta for the ground state. We have used the actual values 
of the energies without any subtractions or rescaling. The last column of Table \ref{VarE-Ph-Pf-MR} shows the results for both PH-PF and MR for 12 electrons.  
A more telling picture of the gaps as a function of inverse size is  shown in \
Fig.\ref{fig:Gaps-MR-PHPf} for 8, 10 and 12-electron systems. 
The stark difference in the gaps between the MR-Pf the and PH-Pf is clearly visible.

In summary, we have presented an exact method for projecting the PH-Pfaffian as well as its 
quasiparticle states to the lowest Landau level. 
The calculations can be organized in a way that allows efficient use of massively
parallel machines.
We obtained wavefunctions for up to 14 and 12 
electrons 
for the GS and charge excitations respectively. By extrapolating finite-size 
results to large sizes in a pure system, we unequivocally find that the PH-Pf energetically 
falls short 
of the Moore-Read Pf (or aPf) state. Other factors such as 
Landau-level mixing or disorder are unlikely to reverse  these trends.

\begin{acknowledgments}
We thank Steve Simon, Mike Zaletel, Zlatko Papic, and Jie Wang for helpful discussions. 
The authors gratefully acknowledge DOE support under the grant \protect{DE-SC0002140}.
\end{acknowledgments}

\bibliography{PH-Pf}

\end{document}